\documentclass{llncs}

    \usepackage{graphicx}
    \usepackage{wrapfig}
\usepackage{amssymb}       %%% mathematical symbols
\usepackage[latin1]{inputenc}
\usepackage[english]{babel}
\usepackage[T1]{fontenc}

    \usepackage[a4paper=true,  bookmarks=false, linkcolor=black, citecolor=black,
      urlcolor=black,colorlinks=true,pagecolor=black, breaklinks=true, bookmarksopen=false,
      pdftitle={Escaping the Trap of too Precise Topic Queries},
      pdfauthor={Paul Libbrecht},
      pdfkeywords={searching for mathematical documents, topics, definitions, web, user interface, learning resources, classifications, subjects},
     ]{hyperref}

\begin{document}
\title{Escaping the Trap of too Precise Topic Queries\thanks{The final publication is available at http://link.springer.com.}}
\titlerunning{Escaping the Trap of too Precise Topic Queries}  % abbreviated title (for running head)
  \author{Paul Libbrecht
     \institute{Center for Educational Research in Mathematics And Technology (CERMAT), Martin Luther University Halle-Wittenberg, 
     Germany\\
     \url{http://www.cermat.org/}
     }}
\maketitle              % typeset the title of the contribution

\begin{abstract}
At the very center of digital mathematics libraries lie controlled vocabularies which qualify the {\it topic} of the documents. These topics are used when submitting a document to a digital mathematics library and to perform searches in a library. The latter are refined by the use of these topics as they allow a precise classification of the mathematics area this document addresses. However, there is a major risk that users employ too precise topics to specify their queries: they may be employing a topic that is only "close-by" but missing to match the right resource. We call this the {\it topic trap}. Indeed, since 2009, this issue has appeared frequently on the i2geo.net platform. Other mathematics portals experience the same phenomenon.  
An approach to solve this issue is to introduce tolerance in the way queries are understood by the user. In particular, the approach of including fuzzy matches but this introduces noise which may prevent the user of understanding the function of the search engine.

In this paper, we propose a way to escape the topic trap by employing the navigation between related topics and the count of search results for each topic. This supports the user in that search for close-by topics is a click away from a previous search. This approach was realized with the i2geo search engine and is described in detail where the relation of being {\it related} is computed by employing textual analysis of the definitions of the concepts fetched from the Wikipedia encyclopedia.

\end{abstract}

% =============================================================
\section{Searching by Mathematical Topics}
% =============================================================

The problem to retrieve mathematical documents in large collections becomes an everyday challenge with the ever-growing collection of digital texts that mathematics professionals and learners have access to.

Word based search in mathematical documents, while still prevailing, suffers from broad issues: formul{\ae} are not covered and, more importantly, the same concepts may be expressed in different terms or with the same terms but in a different sentence organization. For example, the concept of {\it right angle} is often expressed in sentences such as {\it the angle in $A$ is right}, or as $\alpha=90^{\circ}$. 

To alleviate this issue, classifications have been established and mathematical documents annotated by them. The document is {\it indexed} by them. They can be called topics or subjects, even though philosophy of science would treat these concepts as fundamentally different, we consider these names to be equivalent in this paper. Nowadays, most research papers in mathematics are annotated using the {\it Mathematics Subjects Classification (MSC)}\footnote{The Mathematics Subjects Classification is the most commonly used classification for mathematical documents at the research level. It can be browsed through its main catalog at \url{http://msc2010.org/}. A description of recent developments is given in~\cite{MSC2010-Ion-Sperber-EMIS2012}.} and platforms to offer search with them are available. In education, several repositories have been created to share learning resources in communities of practice: the contributed resources are annotated with the topics being covered, as well as many other properties (rights, typical age, instructionnal function, ...). All offer users to search and contribute using topics.

Topic based engines employ a classification of topics. They allow users to search not only by words (and maybe by formul{\ae}) but by the concepts or scientific domains that the resource treats. Examples of such classifications include the MSC, the EUN LRE Thesaurus,\footnote{The EUN LRE Thesaurus is a classification of learning material topics aimed at comprehensiveness across Europe but not at a very deep precision. It is available in 15 languages and contains, for mathematics, less than 20 topics. See \url{http://lreforschools.eun.org/web/guest/lre-thesaurus} for its list.} and the GeoSkills ontology (described in Section~\ref{sec:i2geo-search}).

The usage of such classifications effectively diminishes the ambiguity of the search by words in the text as it diminishes the choices of possible expressions. Moreover, the classifications can, generally, be expressed in multiple languages and thereby allow to find search results in multiple languages, even if the user does not understand them. We shall survey below different methods of choosing the topic.

Even though topic based search can be an effective way to drill down the number of documents matching a word, it also strongly depends on the chosen topic and users may find themselves very quickly {\it trapped in the niche} of a topic that is too precise. Innovative ways to relax the queries when facing an insufficient search result are the focus of this paper.

\bigskip
In the experience of the author, including reports of multiple users using the i2geo platform, searching by topic is permanently compared to searching by words: users will often attempt and mistrust a topic that has been entered if they realize that some documents may be about that topic but have not matched the search. Doing so, they quickly realize that word based search suffers from multiple issues (sketched, among others, in~\cite{LibbrechtKortenkampMercat-IntergeoPlatform-DML09}). 
Reports such as the log-books analyzed in~\cite{Libbrecht-CERME-2013-Role-of-Metadata} indicate this hesitation.

With word searches, at least three solutions exist to offer tolerant or fuzzy matching:
\begin{description}
\item[Partial Results] In almost all search engines as well as i2geo.net the search method assumes that, by default, most of the ingredients of a query (for example the different words) are not queried as a conjunction but as a weighted disjunction: show first the results that match all words, but include also those that only contain single words. This common practice often fails at informing the user about the quality of a series of match and thus introduces {\it noise}.
\item[Latent Semantics] Another way, which has been the basis of~\cite{landauer1997-LSA} is to employ a vector space model of the word/document occurrence and deduce a distance between documents from distance in these spaces. It has been caled latent semantic analysis (LSA). Given a relatively homogenous corpus of texts, this provides effective ways to detect relationships between words and between documents and thus allows search results to include documents that match terms {\it nearby}. Little research has investigated this approach for mathematics (we know of~\cite{Cairns-Informalising-Search} alone).
\item[Suggested Terms] Another way, commonly practiced in contemporary web search engines, is the suggestion of terms that are likely to match the user's query while the original one would lead to no search result as described in~\cite[\S 4.3]{SearchUIs-Hearst}. A version of this feature is the widespread Google search engine's ``did you mean'' as well as its suggested queries. Thus far, little query suggestion has been used for mathematics search engines.
\end{description}

% ----------------------------------------------
\subsection{Outline}
% ----------------------------------------------
This paper first draws a panorama of the user interfaces that search by topics. It then describes the search mechanism of the i2geo.net platform, followed by the core contribution: a method to suggest {\it related} topics having searched for a topic and its implementation. A discussion sets the contribution in perspective. As conclusion, we sketch future works.

%\ednote{Reviewer 3: -Some of the ideas introduced in this paper may be better explained using conventional information retrieval (IR) framework. Some examples are:
%-- (multi)faceted search : seems to be approximately corresponds to topic-based search
%-- query expansion : generalizing user?s query using some dictionaries, ontologies, or other knowledge resources
%-- relevance feedback: a query refinement technique based on user?s feedback
%}

%\ednote{Reviewer 1: The author proposes two mechanisms to help users of i2geo.net locate curricular resources of interest by identifying positions in the GeoSkills ontology that match a keyword query: automatically identifying related positions that should be investigated and displaying the number of resources that match each position to give a sense of the volume of materials available at that position. The first of these is a form of query expansion based on ontologies and the second is a form of relevance feedback.}

% =============================================================
\section{Selecting Topics and Searching}
% =============================================================
\label{sec:choosing-topics}
The choice of the right topic in a search process is a key step. While seasoned mathematicians will often only search the topics of the communities they are used to, commonly calling them by number in the case of the MSC, there are multiple usage types where the user is not necessarily aware of the complete classification. They include searches for literature by a mathematician outside his or her domain of expertise, searches for literature by non-researchers, and searches for learning resources by newcomers in the learning resources platform.

An example is the concept of complexity of an algorithm. It can be found in 7 domains of the MSC, including \textsf{Mathematical logic and foundations} and \textsf{Computer science}. What would a student enter as classification key? The process of choosing the right topic is often a preparation step that is not done in conjunction with the search engine.

Several user interfaces paradigms exist to select a topic:

\begin{wrapfigure}{r}{45mm}\vspace{-0mm}\includegraphics[width=45mm]{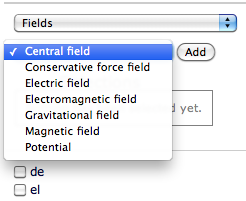}\vspace{-3mm}\caption{Selecting a topic in two steps in the Cosmos portal.}\vspace{-7mm}\end{wrapfigure}
One way to do so is to employ a multi-stage hierarchy where the user first selects a broad topic then finer grained ones. This approach suffers from one major drawback: the hierarchy must be somewhat natural or be learned. As an example, \cite{Megalou-EduTube-Experiment-Draft} reports incomprehension of most pupils when they are told that the \textsf{history} branch is part of \textsf{humanities} when using the EduTubePlus educational video repository\footnote{The EduTubePlus repository ran during the EU project of the same name. See \url{http://www.edutubeplus.info/}. Successor repositories are being built.} browsing by the topics of the EUN-LRE thesaurus. Similarly, the Cosmos portal\footnote{The Cosmos portal is a learning objects repository to share resources pertaining to astronomy in classroom. \url{http://portal.discoverthecosmos.eu/}} has a fine grained classification of physical topics which avoids repetitions; a normal user, thus, needs to search multiple times to find the classical topic of \textsf{optics} which has been put under the main topic of \textsf{waves} while it could have been under the main topic of \textsf{light} or \textsf{fields}.

An enriched version of these approaches is the facetted navigation approach described in~\cite[\S 8.6]{SearchUIs-Hearst}. This approach is richer because it indicates to the user the amount of matching documents before refining or generalizing a query along a hierarchy. In this research we leverage this witness of the query total as an important support to inform the user before choosing a query. However, we claim that navigating up (generalizing) or down (specializing) along a hierarchy may be too restricted and that the user may need alternative ways to navigate to related topics.

Another method of displaying topics to be chosen is by offering {\it tag clouds} which lay out the topics so as to fill the plane by attributing a size to a topic dependent on its frequency in the collection being searched. This method works only well for a small number of topics which are not too inhomogeneous.

Finally, another method to access the topics is that adopted by both major mathematical search engines Zentralblatt\footnote{Zentralblatt Math is a service giving access to abstracts of most of the current and past mathematics research. It is accessible at \url{http://www.zentralblatt-math.org}.} and MathSciNet\footnote{MathSciNet, also called Math Reviews, is a also service to crawl through abstracts of most of the current and past mathematics research. It is accessible at \url{http://www.ams.org/mathscinet/}.}: they suppose that the user will employ {\it subject codes} of the MSC to denote the topics. Both of these search engines do not offer a way to search for the topic, they suppose the codes to be known and let users use other tools like the MSC's main catalog to identify the codes. The euDML library varies this by combining browsing for content and browsing for topics.\footnote{The euDML library gives access to a considerable amount of free access mathematics research papers. It can be browsed by topics at \url{https://eudml.org/subjects/MSC}.} This approach makes it natural for users to switch between topics of the same parent.

The approach employed by the i2geo.net platform is a mix of the approaches above. It employs the search paradigm to allow the user to find the right topic (this is similar to what a newbie can achieve by browsing MSC's catalog), but also allows other forms of navigation by displaying the topic behind a hyperlink everytime the topic is mentioned. This allows users to meet the topic in other places and search for it. 

\begin{figure}[bt]\includegraphics[width=120mm]{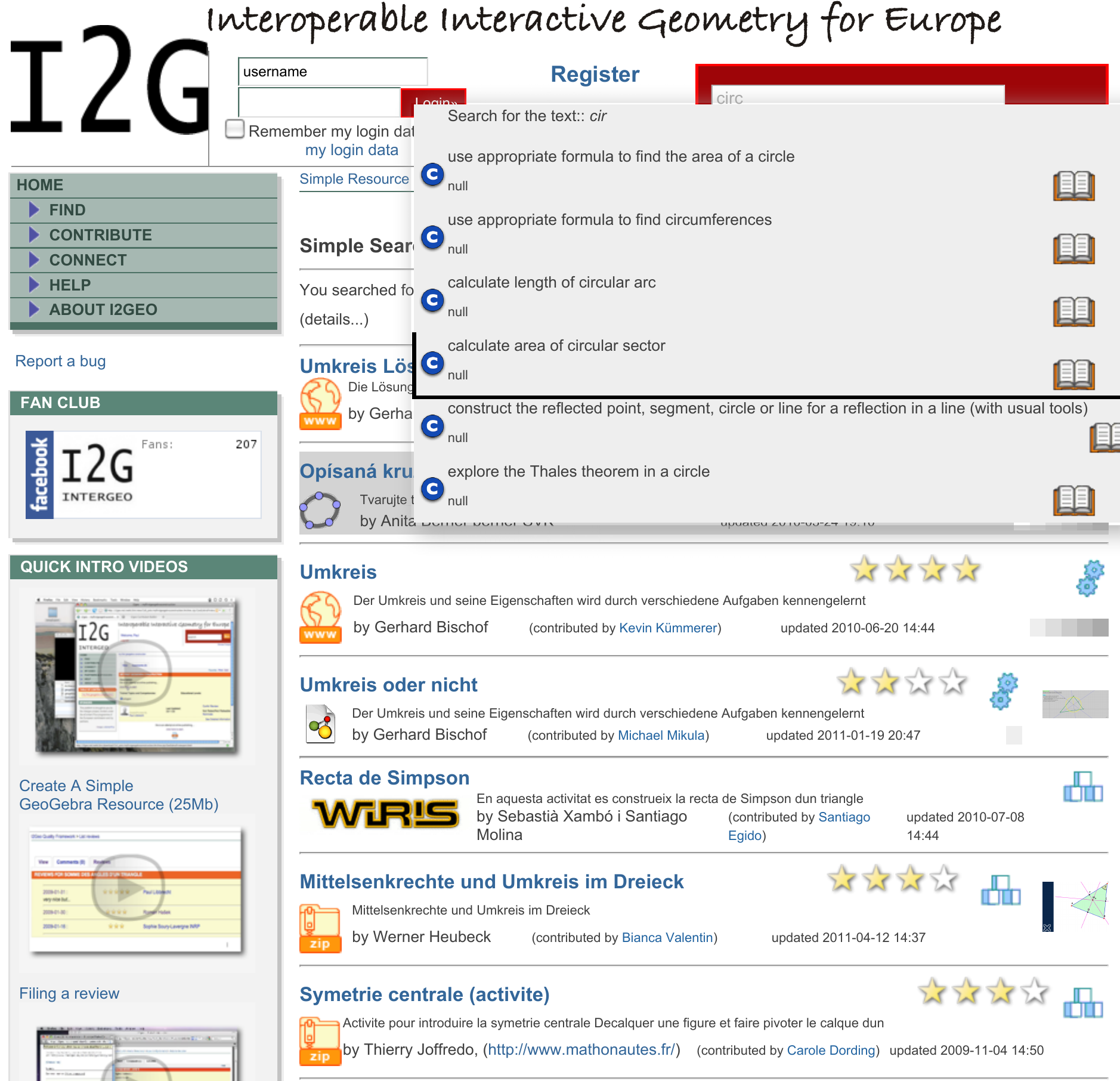}\caption{Searching for circumcircle.}\label{pic:circumcircle}\end{figure}

The approach of i2geo.net aims at being easy to use by teachers of varying proficiency in mathematical science and in the usage of the computer tools. It has, thus, been kept on a single platform with a somewhat consistent user interface. 

The auto-completion phase, where the user searches for a topic, competency, or educational level, is depicted in the Ficture~\ref{pic:circumcircle} where the user searches for the concept of \textsf{circumcircle}, choosing it from the possible choices and find it in several languages below.

In all these approaches, the niche trap of a user stuck in searching for a topic that is too precise is wide open. In the MSC based search above, it is very likely that a user would not attempt to query all the topics pertaining to the complexity; if also including a keyword such as {\it graph} to get informed about complexity of graph algorithms, he or she will either get too many results or too little. Similarly in i2geo, it has been easy to choose a topic that yields zero results and thus fall into an annoying trap.

% =============================================================
\section{The Cross-Curriculum Search of i2geo}
% =============================================================
\label{sec:i2geo-search}

%\ednote{Reviewer 2: chapter 3 seems to be more concerned with promoting i2geo (cross-curriculum search?) than with the topic at hand (I do understand that i2geo has to be introduced to show and understand the new interface, but the focus distribution in this chapter is wrong).}
The i2geo.net platform, described in \cite{LibbrechtKortenkampMercat-IntergeoPlatform-DML09} and available at \url{http://i2geo.net/}, is a learning object repository where teachers of Europe come to share learning resources to learn using dynamic geometry tools. The resources include simple documents of the dynamic geometry systems as well as learning material to support such a learning (exercise sheets, teacher advice, ...). The platform has been built during the Inter2geo project\footnote{More information about the Inter2Geo eContentPlus project can be read in \url{http://i2geo.net/Main/About}.} gathering mathematics education experts from France, Spain, Germany, the Netherlands, and the Czech Republic until 2010. It is now maintained by the University of Halle.

The i2geo search supports search by word and by topic. The topics represent concepts, capacities (which are called competencies), and educational levels of the compulsory mathematics education, they are encoded in an ontology called GeoSkills, each being expressed in the languages of the countries above. The multilinguality of this ontology and the topic based search offered in i2geo supports the mainly graphical nature of many learning resources using dynamic geometry: it is easy to translate a resource of dynamic geometry from one resource to another. The i2geo search has been called a {\it cross-curriculum search}.

The main method of searching is by the choice of a topic, competency, or educational level, and obtaining the matching search results which obeys the following rules which form the ontology based query-expansion mechanism of i2geo search:

\begin{itemize}
\item If a {\it topic} is chosen, i.e. an abstract mathematical concept, resources that have been annotated with this topic are returned and, less preferred, resources that have been annotated by more specific topics.
\item If a {\it competency} is chosen, i.e. a capacity including a verb and topics (see~\cite{SWELBook-GeoSkills}), resources that have been annoated with this competency are returned first and, if not, resources annotated with one of the ingredient of this topic.
\item If a {\it level} is chosen, only resources annotated with that level are returned.
\end{itemize}

The resulting search is effective to find learning resources as soon as the topic is correctly identified but its usage in the last three years has been frustrating for users which often fall in the niche trap of a topic that is too fine as reported, for example, in the log-books of teacher's usage found in~\cite{Libbrecht-CERME-2013-Role-of-Metadata}.

One of the particularly annoying situation has been that of the search competencies which are expected to describe precise capacities that a learner is epxected to acquire. These are typically narrowly defined so that they correspond to a rather isolated place in the curriculum standard. Thus searching for a competency is a very precise action, so much that it returned zero results in most of the cases. Thus, we have made it deliberately tolerant: it should not only return the resources annotated with that competency but resources with are matching any of its ingredients; for example, the comptency \textsf{use the intercept theorem to calculate magnitudes}\footnote{This competency can be browsed at~\url{http://i2geo.net/comped/showCompetency.html?uri=use_intercepttheorem_to_obtain_magnitudes}.} has as ingredients, the competency process \textsf{use in calculating magnitudes} and the topics \textsf{intercept-theorem}, \textsf{rational number}, \textsf{measure}, and \textsf{proportional}. Even if a {\it highlighting} method would be used (as in~\cite[\S 5.5]{SearchUIs-Hearst}), the user would still only understand the relationship to the query if he would completely understand the ingredients of the competency.

One of the actions that users can make, if they realize that they are lost in a too precise topic, is to click on the topic which is displayed and see its hierarchy, then click {\it add this topic} so that resources matching this parent or this child are displayed.
%\ednote{THINKME: figure for this? e.g. with a competency}
However this operation has been repeteadly evaluated as too heavy, with users switching to words queries. So as to realize a simpler switch, we have developed a query suggestion mechanism which suggests to the user {\it related} queries, including parent and children nodes (for concepts), referenced concepts and competency process (for competencies), and related concepts which we describe in the following section.

\begin{figure}
%\vspace{-0.5cm}
\includegraphics[width=140mm]{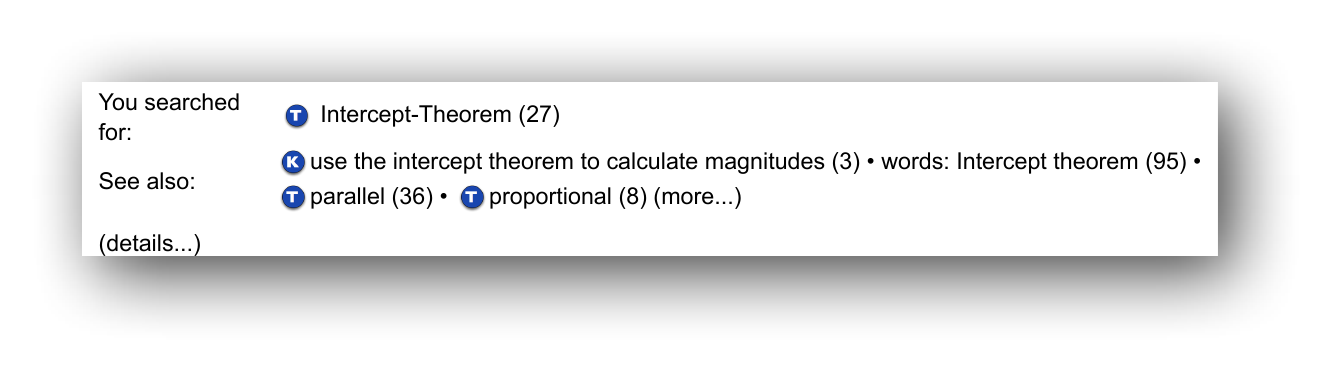}
\vspace{-1cm}
\caption{Suggested queries related to the concept of \textsf{intercept theorem}.}
\label{fig:search-suggestions}
\end{figure}

Moreover, the search index of the terms (topics, competencies, levels) has been enriched with counts of matching resources. This allows the count to be displayed everytime the term is displayed.
This inclusion allows the input of topics to be restricted to those that would not yield empty search results. The query suggestions are depicted in figure~\ref{fig:search-suggestions}.

% THINKME: xxx competency precise and fuzzy as well?? (really needs implementation)

% =============================================================
\section{Computing Related Topics through LSA and Popular Definitions}
% =============================================================
%\ednote{Review 2:Chapter 1 is ok, the presentation of the word-search ideas is too long, I don't even see the relevance of the first solution there at all?}
In order for related topics to be computed, an ontology was created which bases on the GeoSkills' ontology and adds the property of {\it being related} as a reified relationship between the nodes of the GeoSkills ontology. Such a relationship is enriched with a {\it similarity} between $0$ and $1$.

\begin{itemize}
\item As a first step, a part of the relationships are authored manually using Protégé OWL 4\footnote{Protégé is a widespread ontology editor. See \url{http://protege.stanford.edu/}. It has been used for most ontology engineering tasks around GeoSkills in versions 3.} to state the important relationships between concepts that are likely not to be achieved by the text similarity methods below.
Such a relationship exists, for example, to connect the topic of \textsf{circular diagram} and of \textsf{pie chart} which are two concepts that are very common in statistics education. They almost have the same semantics but they have not been clearly identified as equivalent with certainty (several experts have hypothesized that circular diagrams allow chords that are not diagrams to be allowed as limits of the domains). 
These different topics have been contributed by mathematics education experts, mostly teachers, so we contend that such a differentiation of topics is useful even though it is subtle. Coping with such a differentiation without disorienting is, thus, part of the mission of the i2geo search engine and the query suggestion mechanism is a step forward: these two concepts are flagged as close relatives (with the biggest similarity).
\item As a second step, relationships are simply built by policy: parent concepts, competency ingredients, educational levels, age, or educational regions.
\item As a third step, relationships are built by analogy between definitions. To this effect, concepts have been annotated with the URLs to the sections of a Wikipedia page stating a definition of it, for each of the language. Currently this annotation has been done for the languages the author masters, English, French, and German. In each language, a text analysis process, similar to the Latent Semantic Analysis~\cite{landauer1997-LSA}, is performed on these definition texts which can compute similarity distance between the concepts. The SemanticVectors library is used for this~\cite{widdows-semantic-vectors}.
\end{itemize}

Because it is built on different corpora for different languages, this suggestion mechanism takes in account the semantic fields of each language. For example, the fact that a \textsf{disc} is the surface inside of a circle in English and French makes it close to other regions of the plane while the closest German translation of the mathematical disc, {``eine Scheibe''}, is a cylinder, with a thickness. While a strict organization might have decided that {``Scheibe''} is not the appropriate translation, other namings that attempt to denote the interior of a circle ({``Kreisfläche''}, {``Kreisscheibe''}) in German seem to have never been widespread.

%One of the remarkable results of this research is that the difference in semantic fields between mathematical concepts is widely visible in the distance between concepts. In Figure~\ref{fig:similarity-in-three-languages} are the latent-semantic-analysis based distances between disc in French, German, and English.
%
%\begin{figure}
%TODO3
%\caption{Concepts that are LSA-similar to the concept of {\it disc} in French, English and German}
%\label{fig:similarity-in-three-languages}
%\end{figure}

% =============================================================
\section{i2geo Implementation}
% =============================================================

The search engine of i2geo is based on its sharing platform repository, which stores learning resources and displays them, supporting a complete learning objects sharing mechanism described in~\cite{LibbrechtKortenkampMercat-IntergeoPlatform-DML09} which derives from the XWiki Collaborative Learning Asset Managements System (XCLAMS\footnote{Technical details on the open-source XCLAMS project can be found \url{http://xclams.xwiki.org}}). The resources search index extends Apache Lucene with i2geo and XWiki information.  This user interface extends the basic text search mechansim with the search weighting described in section~\ref{sec:i2geo-search} and with the chooser for topics, competencies, and levels. This chooser is a JavaScript component which fetches results from the auto-completion index as the user types.

The auto-completion index, also based on Lucene, is populated by reading the ontologies: the GeoSkills ontology~\cite{SWELBook-GeoSkills}, the ontology of relationships (called {\tt GeoSkills-Relatives}), and the ontology of subjects. The ontology is read using the library OWLapi and the reasoner Pellet.

Similarly as the GeoSkills ontology and its displays such as the CompEd web-application~\cite{Desmoulins-Libbrecht-CompEd-SWEL09}, which are designed to be updated on a daily basis to take advantage of the changes by the curriculum experts, the GeoSkills-Relatives ontology is designed to be updated when changes are made either in the definitions' texts or in the manual relationships. While GeoSkills would receive updates from curriculum experts, the GeoSkillsRelatives ontology would receive updates from developers receiving feedback from users and encoding them using Protégé, and from the definition texts fetched from their web-sources, currently Wikipedia. These ontologies are then published and they enter the rebuild of the auto-completion index which, also, queries the numbers of matching resources for each term.

The code of the software and the ontologies are open-source, under the Apache Public License, and available from \url{http://github.com/i2geo/}. It has not yet entered the production site of i2geo.net, as it is being polished.

% =============================================================
\section{Discussion}
% =============================================================

Having described the technical realization of this approach, we now discuss its relevance to research areas which are closely related to it.

\subsection{Navigation Between Topics}
The approach of suggesting queries we have developed supports the user in exploring the topics' organization, inviting them to look at topics which are laterally related to the queried topics. This form of navigation shares the practice of facetted navigation of~\cite[\S 8.6]{SearchUIs-Hearst} in that it suggests queries, but it does so without being restricted to the hierarchy of topics.
Moreover, users can explore the structure of topics: they can click on the queried topic and see a {\it topic page}.
%\ednote{ako: Didn't you call this ``topic view'' before? Personally, I liked topic view better. PL: it does not seem to be spoken out in other places. The problem with view is the relationship to ``point of view'' or ``database view'' which are restrictions of a bigger view; I think this interpretation is wrong}
The topic pages of i2geo, served by the CompEd server, perform a similar role as the MSC catalog: they allow users that wish to inquire about the classifications' structure to navigate through it. 
%\ednote{add picture of a search the topic page on top. E.g \url{http://i2geo.net/comped/showTopic.html?uri=Circumcircle} ?}
Zentralblatt Math and euDML present any occurrence of an MSC subject as a hyperlink to  the query for this topic but do not provide a link to the context, that is, its hierarchy. In contrast, the i2geo search links them to {\it topic pages} (except for the suggested queries which display the number of matching resources and link to the search result).
It is probable that the approach of presenting this as a pop-up window, as i2geo does it, is inappropriate for contemporary web habits. However the topic pages provide information to users.
The MSC Catalog and euDML search engine also render the topic in its environment wrt other topics; however no further information is provided about each of the subjects.

Should such links be such as euDML? As a drill-down of the search? As a topic page?
Topic pages could include illustrations, or links to a Wikipedia or MathWorld, which would draw on users visual memory, so that they are supported in recognizing topics and thus better remember them. 

Alternatively, a few attempts have been made to navigate through mathematical topics by the display of a graph that represents a map of the topics' relationships. Examples include the server \url{http://thesaurus.maths.org/} (now closed), the knowledge map of the Khan Academy's Knowledge Map (\url{https://www.khanacademy.org/exercisedashboard}), or the MSC map at \url{http://map.mathweb.org}. However, all these approaches consume a large screen space and are thus difficult to combine with the search activity.

It remains open if additional navigation mechanisms, be them topic pages or graphical knowledge maps, are effective in providing the users of the search tool an awareness of the structure and nature of the topics so that they can be chosen effectively.

\subsection{Applicability for MSC Subjects}
What are the challenges to port the approach we described to the subjects of the Mathematics Subjects Classification? It relies on a few ingredients: the ontological structure between topics, the multiple names of each so as to allow search and identifiable display, and the existence of Wikipedia pages to describe them so that additional relatedness connections are computed. 

One of the challenges
%\ednote{ako: add ``though'' PL: this does not oppose to anything, it is a challenge of applying the approach. Changed ``how difficult'' to `` what are challenges'' above} 
lies in the user interface: when unknown to the user, subjects are best understood when displayed within the hierarchy as this provides context information. For example, the subject  \textsf{14Q20 Effectivity, complexity} if suggested, has much chances to confuse users, since its belonging to algebraic geometry is not explained. Similarly the subject \textsf{97H60 Linear algebra} when displayed would miss its belonging to the subject \textsf{97-XX Mathematics education}.

Another challenge lies in finding pages that form a source of descriptions of subjects to perform the computation of relatedness between the subjects. While pages in Wikipedia exist for main subjects, they are largely missing for more refined subjects.

\subsection{Mix and Match on the World Wide Web}
\label{sec:mix-match-www}
More and more markup formats are available to allow web-page authors to indicate inside web-pages such properties as {\it being of a given topic}. The anchoring of topics within a structure such as a taxonomy or ontology makes them a particularly good source for identifiable keys to describe the topics.

Among the main markups that allow this indication is the family of microdata annotations at \url{http://schema.org/}, in particular {\tt CreativeWork}'s {\tt about} property. This property can carry both a text and a URI so that such initiatives as the standardized thesaurus encoding of MSC~\cite{MSC2010-Ion-Sperber-EMIS2012} in the SKOS language or the ontological nature of GeoSkills in the OWL language can be easily encoded there.

It appears, from schema.org's intro pages, that such an inclusion in  web-pages of digital mathematics libraries would allow the main web search engine to give a significant weight to the topics annotations and be able to exploit the classification structure. Thus they could suggest generalizations to the domain of \textsf{category theory} when mentioning \textsf{topoi}, or to \textsf{perpendicular bisector} when mentioning \textsf{circumscribed triangle}. This could complement the suggested queries offered by these systems which are mostly based on the of earlier users' queries and are, thus, almost always absent when inputting refined mathematical topics. 

We have observed that regular users commonly use main web search engine in parallel to the search engine of the digital libraries. We expect that the search engines' complementary features are likely to enrich each others, where users are able to depict features of one to describe desired features of others.

% =============================================================
\section{Conclusion}
% =============================================================
In this paper, we have described a way to enrich the search engines that employ topics so as to avoid the trap of a too precise topic. This trap seems to be common to all search engines that offer this function, including the exemplary facetted search system of~\cite[\S 8.6]{SearchUIs-Hearst}.

The solution we propose is to enable the user in choosing topics that are closely related to the query by the presentation of suggested topics decorated by the number of matches.  For example, it allows a user who has search results about \textsf{ellipse} to go in one click to the search results for \textsf{conic sections}, to \textsf{cone}, \textsf{disc}, or \textsf{meridian}.

The mathematics learning resources' sharing platform \url{http://i2geo.net} has integrated this suggestion mechanism. This integration supports the user to explore other search queries that may satisfy better their expectactions. Anchoring this choice in the display of the available data appears to be an important step to guide the user while still avoiding the pages with empty search results.

Users typically lack the knowledge of the topics classifications employed by search engines. The suggestion mechanism allow them to explore related topics.
% The main justifications of letting the user navigate through the various nodes of the query terms is the lack of knowledge of the classification. 
This lack of knowledge is stated quite clearly by the  teachers trying to find the history domain but needing to navigate through social studies first (see Section~\ref{sec:choosing-topics}). It is also echoed by multiple i2geo users who prefer to switch to text search. 

It could be suggested that such a lack of knowledge could be alleviated by designing a more natural hierarchy or by educating the users in the usages of widespread classifications. This would be quite artificial: many hierarchy decisions are natural in one culture and not in others; this is the case for the history domain which is a part of social studies in the American education systems but not in most European ones. Thus, offering a tool to stimulate the users in exploring the related topics is an important way to support the user into gaining a better knowledge of how topics are navigated and to let them come out of the trap of a too precise topic.

\subsection{Future Work}

Studies currently planned ahead of this research include the following:

A stabilization of the server code is the closest objective. The current implementation, requiring delicate versions of each of the software libraries, has prevented a full deployment to the server. This implementation will reach the users of the \url{http://i2geo.net} platform which is used regularly by users of Europe and beyond (since last year, a mean of 90 search queries per day has been observed).

Another facet is a broader coverage of definition URLs of topics. On the one hand, a fairly modest count of topics has been enriched with the URL of Wikipedia definitions, while a broad part has been automatically guessed without validation. On the other hand, only three languages have been taken up, French, English, and German. We intend to employ the Curriculum-encoders' voluntaries community to organize such contributions (\url{http://i2geo.net/xwiki/bin/view/Group_Curriculum-Encoders/}).

These two development works are the basis to get a usable implementation. They are likely to enhance productivity of users that employ topic based search. Early and recent experiments in which pre-service teachers were discovering i2geo.net have confirmed the need for more stimulations to attempt search by other topics. Such discoveries, generally coupled with an introduction in a course, form an important field-trial-like evaluation of the search platform: the little time they allocate to discover the platform and the relative neutrality of  students make them good candidates to judge the quality of a prime-time experiment. 
%As an example, a recent introduction led one of the student in the search for i2geo learning resources that were annotated about the volume of a cube ({\it Würfelvolumen} in German) and to be disappointed by the lack of such a specific topic while {\sf cube} and {\sf volume} are existing topic; related queries suggestions would lead that person to effectively switch to such topics as the volume of a body or the related competencies.

Finally, as indicated in~\cite{MSC2010-Ion-Sperber-EMIS2012}, it is likely that linking across several ta\-xo\-no\-mies will emerge as a common practice. One of the attempts to do so has been done on \url{http://i2geo.net/xwiki/bin/view/Subjects/} which takes a handful of relevant mathematical domains as search subjects and maps them to GeoSkills nodes allowing users to search for dynamic geometry constructions in probability and statistics, for example. The mapping there is created by ontological axiom statements making equivalent the subjects to the union of topics and competencies. Such a mapping is currently being leveraged by a team of the Open University of Cyprus for the subjects of the MathTax taxonomy\footnote{The MathTax taxonomy is in use in parts of the USA's National Sciencce Digital Library. It can be browsed at \url{http://people.uncw.edu/hermanr/MathTax/}.}. It will allow fine-grained subjects search on \url{http://i2geo.net} and on the upcoming \url{http://opendiscoveryspace.eu} and thus become more compatible with the MSC.

Linking across several classifications, either as done in the i2geo subjects or in other ways described in Section~\ref{sec:mix-match-www}, is likely to shed a new light on the suggestions of related terms since the display of them would be less homogenous. At the same time, such a mix is likely to distill more relevant information from the user's culture and context and thus create more relevant search experiences. It has the potential of exploiting the user's locality or earlier search to influence the weight of suggested queries and thus, for example, prefer suggested queries that are in the user's region's curriculum standard, closer to the user's known concepts, or to the user's currently research topics.

\subsection{Acknowledgements}
This research work has been partially funded by the European Commission under the Policy Support Programme. The opinions expressed in this paper are that of the author. The author wishes to thank Yannis Harlambous for seminal discussions, Cyrille Desmoulins for contributing large parts of the ontological engineering approach of the GeoSkills ontology, and the workers of the Inter2geo eContentPlus project for having provided a realistic environment to deploy modern technology. Moreover, he wishes to thank the shepherd of the programme comittee for the detailed and dedicated review work.

\bibliographystyle{alpha}
\bibliography{Flexibilizing-Queries}

\end{document}